

\input harvmac
\noblackbox
\def\efi{\vbox{\sl\centerline{Enrico Fermi Institute,University of Chicago}%
\centerline{5640 Ellis Avenue,Chicago, Illinois 60637 }}}
\def\con{{\rm cons}}

\def\Atwo{A^{(2)}}

\def\Asix{A^{(6)}}
\def\dtwo{d^{(2)}}

\def\dsix{d^{(6)}}
\def\Dtwo{D^{(2)}}

\def\Dsix{D^{(6)}}

\def\Ftwo{F^{(2)}}
\def\Ffour{F^{(4)}}
\def\Fsix{F^{(6)}}
\def\Rsix{R^{(6)}}
\def\omegasix{\omega^{(6)}}
  \def\CC{{\cal C}}

\def\tr{\mathop{\rm tr}}
\def\Tr{\mathop{\rm Tr}}

\def\ie{\hbox{\it i.e.}}

\lref\ch{C. Callan and J. A. Harvey,  Nucl. Phys. {\bf B250} (1985) 427. }
\lref\nacu{S. Naculich, Nucl. Phys. {\bf B296} (1988) 837.}
\lref\kaplan{D. Kaplan, Phys.Lett. {\bf288B} (1992) 342.}
\lref\condm{D. Boyanovsky, E. Dagotto and E. Fradkin,
Nucl. Phys, {\bf B285} (1987) 340 \semi M. Stone and F. Gaitan,
Ann.  Phys. {\bf178} (1987) 89 \semi G. E. Volovik, ``Vortex motion in fermi
superfluids and Callan-Harvey effect'', preprint 1993. }
\lref\gs{M. Green and J. Schwarz,  Phys. Lett. {\bf149B} (1984) 117.}
\lref\klee{ K. Lee, ``The Dual Formulation of Cosmic Strings and Vortices'',
preprint CU-TP-588 (1993).}
\lref\fbranes{see C. Callan, J. A. Harvey, and A. Strominger, ``Supersymmetric
String
Solitons'' in String Theory and Quantum Gravity '91, Ed. J. Harvey, R. Iengo,
K. Narain,
S. Randjbar-Daemi and H. Verlinde, World  Scientific (1992) for a review. }
\lref\hetsol{A. Strominger,  Nucl. Phys. {\bf B343} (1990) 167.}
\lref\wz{J. Wess and B. Zumino, Phys. Lett. {\bf B37}  (1971) 95. }
\lref\bz{W. A. Bardeen and B. Zumino, Nucl. Phys. {\bf B244} (1984) 421.}
\lref\iz{J. M. Izquierdo and P. K. Townsend, preprint DAMTP/R-93/18.}
\lref\hw{C.  M. Hull and E. Witten, Phys. Lett. {\bf160B} (1985) 398.}
\lref\witt{E. Witten, Phys. Lett. {\bf153B}  (1985) 243.}
\lref\dh{A. Dabholkar and J. A. Harvey, Phys. Rev. Lett. {\bf 63} (1989) 719.}
\lref\dghr{A. Dabholkar, G. Gibbons, J. A. Harvey  and F. Ruiz Ruiz, Nucl.
Phys.
 {\bf B340} (1990) 33.}
\lref\ddp{J. A. Dixon, M. J. Duff and J. C Plefka, Phys. Rev. Lett. {\bf 69}
(1992) 3009.}
\lref\heta{D. J. Gross, J. A. Harvey, E. Martinec and R. Rohm, Nucl. Phys. {\bf
B256} (1985) 253.}
\lref\duffstr{M. J. Duff and J. X. Lu, Phys. Rev. Lett. {\bf 66} (1991) 1402.}
\lref\bcgw{C. W. Bernard, N. H. Christ, A. H. Guth and E. Weinberg, Phys. Rev.
{\bf D16} (1977)
2967.}
%
\Title{\vbox{\baselineskip12pt
\hbox{EFI-93-55}
\hbox{hep-th/9310035}}}
{Anomaly Inflow for Gauge Defects }
{

\bigskip
\centerline{Julie Blum and Jeffrey A. Harvey }
\bigskip
\efi
\centerline{\it Internet: julie@yukawa.uchicago.edu}
\centerline{\it Internet: harvey@poincare.uchicago.edu}

\bigskip
\medskip
\centerline{\bf Abstract}

Topological defects constructed out of scalar fields and possessing
chiral fermion zero modes are known to exhibit an anomaly inflow
mechanism which cancels the anomaly in the effective theory of
the zero modes through an inflow of current from the space in which the
defect is embedded.
We investigate the analog of this mechanism for defects constructed
out of gauge fields in higher dimensions.  In particular we analyze
this mechanism for string (one-brane) defects in six dimensions
and for fivebranes in ten dimensions.}

\Date{10/93}

\newsec{Introduction}
In earlier work the gauge
and gravitational anomalies of chiral fermion zero modes bound to
scalar defects such as axion strings or domain walls were
investigated \ch.  It was shown
that these anomalies are physically sensible when the embedding
of the defect in a
higher dimensional space and the possibility of charge inflow
from this higher-dimensional space is  included. Although
fermions outside the defect are massive, they nonetheless induce
a vacuum current outside the defect in the presence of external
gauge fields. This current is conserved away from the defect
and has a divergence on the defect which precisely cancels
the anomaly present in the low-energy effective action of the
chiral zero modes bound to the defect.
In addition,  the whole process occurs in a way which
is consistent with overall gauge covariance \nacu. This mechanism
provides a simple physical model of the mathematical relation between
chiral anomalies in $2n+2$ dimensions and non-abelian (or gravitational)
anomalies in $2n$ dimensions.   There have also been recent discussions
of the application of this effect both in condensed matter physics \condm\
and in  theories of chiral lattice fermions \kaplan.

This paper generalizes the analysis of
\ch\ and \nacu\  to gauge defects (constructed from Yang-Mills
instantons) in $4k+2$
dimensions.  For scalar defects the starting point was a consistent
anomaly free theory with a chiral coupling of the  fermion fields  to
the scalar fields responsible
for the defect. The current inflow could then be
deduced from the presence of couplings  of the defect to gauge fields
which could be viewed as the result of integrating out the massive fermion
degrees of freedom. In trying to extend this picture to gauge defects
there is one main new complication. In order to obtain chiral fermion
zero modes it is necessary for the gauge fields to have chiral couplings
to the fermions. But,  in dimensions greater than four, such chiral gauge
theories are generically inconsistent because of gauge
anomalies.  A similar problem arises if we wish to consider energy-momentum
inflow and chiral gravitational couplings.
However in certain cases of physical interest it is known
that the gauge and/or gravitational anomalies can be cancelled through
the Green-Schwarz mechanism \gs. This involves introducing a two form
field $B$ which has non-trivial transformation properties under gauge and
local Lorentz transformations and certain higher dimensional couplings of
$B$ to the gauge and gravitational fields.  These anomaly cancelling terms,
at least in the context of string theory, can be viewed as terms which
arise in the low-energy effective Lagrangian as a result of integrating
out the massive modes of the string. Thus these theories with anomaly
cancellation are consistent gauge invariant theories which can have gauge
defect solutions with chiral fermion zero modes. By analogy with the
earlier studies of scalar defects one  would expect  the apparent
anomaly in the effective theory of the chiral zero modes to be canceled
by  an inflow from outside the defect. One  might further expect
the anomaly canceling
terms to play an important role in understanding how this comes about.
As we will show, these expectations are correct.

The outline of the paper is as follows. In the second section
we review the inflow mechanism of  \refs{\ch, \nacu}  for axion strings in four
dimensions.
In the third  section we introduce  a simple model in six
dimensions which has a string solution constructed using
Yang-Mills instantons and in which the fermion gauge
anomaly is cancelled via the Green-Schwarz mechanism.
In this model we then show how the
gauge anomalies in the low-energy effective action for the
chiral zero modes are cancelled by inflow from the outside
world. In the fourth section we extend this analysis to fivebrane
solutions of the low-energy limit of heterotic string theory which are also
constructed using Yang-Mills instantons.
We analyze both the gauge and gravitational anomalies
on the fivebrane and show how they are cancelled.
The final section briefly discusses some of the implications
of these results for fundamental and solitonic strings and fivebranes.
While this work was in progress we received a paper by
Izquierdo and Townsend \iz\  which addresses some of the same
issues as this paper.  We will comment on their work in the
final section.

\newsec{Anomaly inflow for axion strings}
In this section we will briefly review the inflow mechanism of  \refs{\ch,
\nacu} as it
applies to axion strings in four dimensions.  We consider a complex
scalar field $\Phi = \Phi_1 + i  \Phi_2$ with a non-zero vacuum
expectation
value $v$. An axion string configuration  with winding number $n$
is given by
\eqn\one{ \Phi = f(\rho) e^{i \theta}}
with $\theta = n \phi$
where $(\rho, \phi)$ are polar coordinates in the plane transverse
to the string and where $f$ goes to zero at the origin and to $v$
at infinity.

We have in addition Dirac fermion fields $\psi$ in some representation
$\bf r$ of a gauge group $G$  which interact
both with $\Phi$ and with the gauge field $A_\mu$
\eqn\two{ {\cal L} = \bar \psi i \Dsl  \psi - \bar \psi ( \Phi_1 + i \gamma^5
\Phi_2)
                   \psi  .}
Since the gauge coupling in \two\ is vectorial there is no anomaly
in the gauge current.  Equivalently, the effective action resulting from
integrating
out the fermion fields,
\eqn\three{ i S_{\rm eff} = \log  {\rm Det} (i \Dsl - f (\rho)e^{i \theta
\gamma_5})}
is invariant under infinitesimal gauge transformations, $\delta A = D \Lambda$
(from here on we will use the language of differential forms with $A$ a
one-form
which is also an anti-hermitian element of the Lie algebra of $G$, $A= A_\mu^A
T^A dx^\mu$,
and $D$ the
covariant exterior derivative).

The basic puzzle arises because the Dirac equation in the axion string
background
possesses  $|n|$ {\it chiral} two-dimensional zero modes bound to the string
with
the
chirality determined
by the sign of $n$.  It would thus seem that the coupling of these chiral
fermion
zero modes to the background gauge fields must be anomalous and lead
to a lack of current conservation.  However
the full theory is clearly not anomalous. The resolution is that the effective
action \three\  has two physically distinct contributions, each one of which
would
separately lead to a violation of current conservation, but which combine to
give
a theory with a conserved and properly covariant current.

The first contribution arises from the $|n|$ chiral fermion zero modes and is
localized on
the string.  The current derived from the two-dimensional action for the
zero modes in the presence of
a two-dimensional  background gauge field $\Atwo$,
\eqn\four{ j^\con=  { \delta S_{2} \over \delta \Atwo} , }
has a covariant divergence given by the consistent anomaly,
\eqn\five{ \Dtwo  j^\con = - { n \over 4 \pi}  \dtwo \Atwo .}
(we are using conventions where the current in $d$ dimensions is a $d-1$ form
whose dual ${}^*j$ has components equal to the usual current $j_\mu$.)

The second contribution arises from vacuum currents induced by the non-zero
fermion
modes  in the presence of the background gauge field and the topologically
non-trivial phase of the axion field.  The induced current can be calculated
directly
at the one-loop level  or it can be inferred by removing the coupling of the
phase of
$\Phi$ to the fermions by performing a chiral rotation and using the chiral
anomaly
\refs{\ch,\nacu}.
If we work in the ``thin string'' approximation where we freeze $f(\rho)$ to
its
vacuum expectation value $v$ the second approach leads to a coupling
\eqn\six{ S_{\theta} = - { 1 \over 8 \pi^2} \int d \theta \omega_3}
with $\omega_3 = \Tr (AF - A^3/3)$.  In the thin string approximation $\theta$
is not well
defined at the origin, but this can be accounted for by noting that
\eqn\seven{\int_{D^2} d^2 \theta = \int_{S^1} d \theta = 2 \pi n }
where the first integral is over a disk in the two dimensions transverse to the
string and the second
is over the $S^1$ boundary of the disk. Thus we may treat $d^2 \theta$ as being
$2 \pi n$ times
a delta function in the two transverse dimensions.
If we now vary $S_{\theta}$ with respect to $A$ we find a current
\eqn\eight{ J_{\theta} = {1 \over 8 \pi^2}  (2 F d \theta - A d^2 \theta)
\equiv
   J_{\infty} + \Delta J . }

To see that the two contributions combine to give a covariant and conserved
current
we integrate the covariant divergence of the four-dimensional current
$J_\theta$
over the two dimensions transverse to the string and show that this when added
to the
covariant divergence of the two-dimensional current gives zero.
 We again consider a background
gauge field $A=\Atwo$ which is tangent to the string world sheet. We can then
write the covariant
derivative as $D = \Dtwo + d_{T}$ with $d_{T}$ the exterior derivative in the
transverse dimensions.  Noting that $\Dtwo J_\infty = d_{T} \Delta J = 0$ we
have
\eqn\nine{\eqalign{ \int_{D^2} D  J_{\theta} =&  \int_{S^1}  J_\infty +
                                                                 \Dtwo
\int_{D_2}   \Delta J \cr
                                                 = & { n \over 2 \pi } \Ftwo -
\Dtwo  \Delta j }}
where $\Delta j =   n  \Atwo/(4 \pi)$.  Therefore
\eqn\ten{ \int_{D^2} D  J_{\theta} +  \Dtwo  j^\con  =  { n \over 2 \pi} \Ftwo
- { n \over 4 \pi}
(\Dtwo \Atwo + \dtwo  \Atwo) = 0 }
and we verify that the total current is conserved. Physically what is going on
is that
the coupling
of the massive fermions to the gauge field and the axion induce
a radial current $J_\infty$ which flows onto the string from infinity. This
current
is covariant, but has a divergence on the string.  This divergence is matched
by the divergence of the current flowing along the string. The current along
the
string has two contributions. The first comes from the fermion zero modes and
is necessarily {\it not} covariant since it comes from variation of a
two-dimensional
action \wz.  The second contribution, $\Delta j$, while it is localized on the
string,  is not derivable from a two-dimensional action  but only from the
four-dimensional action and is precisely the contribution to the current
needed to convert the consistent current into the covariant current \bz.

An alternative treatment which is closely related to the theories we will
discuss in the next two sections involves writing the theory in a dual
formulation with
a three-form field strength $H = v {}^* d \theta$.
The dual formulation is more natural from a geometric point of view
because the two-form $B$ has a local coupling to the string world-sheet
while the zero form $\theta$ does not.
This formulation has
been discussed in \nacu\ and more recently  in \klee\ so we will be brief.
The dual action is written in terms of the two-form $B$ and its
field strength $H = d B - \alpha' \omega^3$ where
 $\omega_3=  \Tr (A  F - (1/3)A^3)$ is the Chern-Simons three-form
and $\alpha'={1 /(16 \pi^2 v)}$.
$H$ is
gauge invariant provided that $B$ transforms under gauge transformations
as $\delta B = \alpha'  \omega_2^1 \equiv  \alpha' \Tr ( \Lambda d A) $. The
action for $B$ is then
\eqn\eleven{ S_B = \int  H {}^* H + 4 \pi  n v \int B {}^*V }
where $V$ is the volume two-form on the string world-sheet.   Varying
$S$ with respect to $A$ we find a current given by
\eqn\twelve{ J =  2 \alpha'  ( 2 F {}^* H  - A d {}^* H) =  2 \alpha'
                       ( 2 F {}^* H - 4 \pi n v A {}^* V)}
where we have also used the $B$ equation of motion $ d {}^* H = 4 \pi n v {}^*
V$.
The first term in \twelve\  agrees with $J_\infty$  while the second term gives
the
correction to the current  $\Delta J$.

\newsec{Six-dimensional Instanton String}
We now turn to a theory which has chiral fermions coupled to gauge fields in
six spacetime dimensions.  We can use the usual four-dimensional Euclidean
Yang-Mills instanton to construct  a one-brane (string) topological defect
in this theory.

Our starting point is the
action  for a gauge field with gauge group $G$ , Weyl fermions
in the representation ${\bf r}$ of $G$, and an antisymmetric tensor field
$B_{\mu \nu}$ which will be necessary for anomaly cancellation.
 \eqn\thirteen {S_0=\int
d^{6}x \left(
  - { 1 \over 4g^2}  \langle F_{\mu\nu}F^{\mu\nu} \rangle+
      {1 \over 2} \overline\lambda(\Dsl \lambda)  - {1 \over 12}
       H_{\mu\nu\rho} H^{\mu\nu\rho} \right) }
The gauge
field strength two-form and antisymmetric tensor field strength three-form
are
\eqn\fourteen{\eqalign{ F & = d A + A^2  \cr
                               H &= d B - \alpha' \omega_3 . }}

With our conventions the gauge coupling $g$ has dimensions of
length and $\alpha'$ is an independent dimensionless coupling. Traces will
always be taken in the fermion representation ${\bf r}$.  The angular
brackets denote the Cartan inner product in the Lie algebra of G
\eqn\fourthalf{ \langle T^A T^B \rangle = - \CC \Tr (T^A T^B) =  \delta^{A B}
.}
The normalization constant $\CC$ will be specified later when we consider
specific examples.
The bosonic
part of the action can be written as
\eqn\fifteen{ S_B = {1 \over 2}  \int ( {1 \over g^2}  \langle F {}^* F \rangle
 + H {}^* H ) }
and is invariant under the infinitesimal gauge transformation
\eqn\sixteen{\eqalign{ \delta_\Lambda A & = D \Lambda \cr
                                 \delta_\Lambda B &= \alpha' \omega_2^1  \cr }}
where $\omega_2^1 = \Tr (\Lambda d A)$  obeys
$\delta_{\Lambda} \omega_3 = d \omega_2^1$.

While this theory is classically gauge invariant, at the one-loop level
is has a
gauge anomaly which can be determined
using the descent equations starting from the chiral anomaly in $8$
dimensions:
 \eqn\seventeen
{I_{8}(F)={1\over(2\pi)^{4}(4!)} \Tr (F^{4}) .}
In order for
the anomaly cancellation mechanism of \gs\ to work it is necessary that the
quartic
trace in \seventeen\  factorize in the form
\eqn\eighteen{\Tr F^4 =c\;  (\Tr(F^2) )^2}
for some constant $c$. The anomaly derived from \seventeen\ by descending to
six
dimensions
is then
\eqn\nineteen{\eqalign{ G_{6}(\Lambda)= & -  \int
\Tr (\Lambda D J_f) \cr
 = &  c' \int w^1_2 \Tr (F^{2}) , \cr }}
where $c' = c/(4! (2\pi)^3 )$ and $J_f$ is the fermion gauge current.

Given the factorization \eighteen\ we can
add a term to the action which cancels the anomaly by utilizing the
non-trivial gauge variation
of $B$
 \eqn\twenty{\Delta S= - { c' \over \alpha'} \int  B \Tr (F^2) . }
The action $S= S_0 +  \Delta S$ then defines a gauge invariant
theory (in this section we will ignore potential gravitational anomalies).

For the axion string it was clear
that the overall gauge current must be conserved since there were
no gauge anomalies.  What is the situation here?  If we integrate
out the fermions, then the resulting effective action, as a function
of $A$ and $B$,  is gauge invariant
by the above construction.  This implies that
\eqn\twfour{ \delta_{\Lambda} S_{eff}  =  \int {\rm Tr} (D \Lambda
                  { \delta S_{eff} \over \delta A } ) + \alpha'  \omega_2^1
{\delta S_{eff}
                  \over \delta B} = 0 . }
Integrating by parts on the first term we see that there is a conserved
current $J = \delta S_{eff} / \delta A$ for backgrounds obeying the
$B$ equation of motion $\delta S_{eff} / \delta B =0$.

The equations of motion which follow from variation of $S$ are
\eqn\twone{\eqalign{ d {}^* H & =  - {c' \over  \alpha'} \Tr  (F F) \cr
        - {1 \over g^2}  D {}^* F & = J_f   + J_b  \cr }}
where $J_f$ is the fermionic contribution to the current and
\eqn\twtwo{ J_b = -2 \alpha' F {}^* H + \alpha' A d {}^* H - {2 c' \over
\alpha'}
                             F dB }
is the bosonic contribution to the current.   Using \nineteen, the first of
\twone, and \twtwo\
we check that  the total current
is covariantly conserved, $D(J_b + J_f) =0$

We now want to analyze current conservation in a background consisting
of an instanton string. This is
a configuration with topology $M^2 \times R^4$ with $M^2$ being two-dimensional
Minkowski space.  We can split the exterior derivative and gauge
one-form into two- and four-dimensional parts as
\eqn\twfive{ d = d^{(2)} + d^{(4)} ; \qquad A = A^{(2)} + A^{(4)} . }
For the gauge fields we take $A^{(4)}$ to be a Yang-Mills instanton,
embedded in a $SU(2)$ subgroup of the gauge group $G$.  In order
to study current inflow we then turn on an additional  gauge field $A^{(2)}$
which lives in a subgroup of $G$ orthogonal to the instanton $SU(2)$.
We will call this  subgroup $G'$.
We assume for simplicity that $A^{(4)}$ and $A^{(2)}$ depend only on the
coordinates on $R^4$ and $M^2$ respectively.

%
%
%
%

As in the axion string case, the fermion contribution to the current
has two parts, one localized on the string and the other flowing in
from infinity.  The new feature is that there is also a bosonic contribution to
the
current coming from the couplings of the antisymmetric tensor field.

We will first analyze the fermionic contribution.  In order to keep the group
theory to a minimum we will first do the analysis for a simple
choice of the gauge group. We take $G=SU(3)$ with the fermions
in the ${\bf 3}$ of $SU(3)$ and set $T^A = -i \lambda^A/2$ with the $\lambda^A$
the usual Gell-Mann matrices. We then have $\CC= 2$ and $c=1/2$. We imbed the
instanton in the minimal $SU(2)$ subgroup $SU(3) \supset SU(2) \times U(1)$
with
\eqn\twfhalf{ {\bf 3} \rightarrow {\bf 2} ({1 \over 2 \sqrt{3}}) + {\bf 1}(-{1
\over \sqrt{3}}).  }

An instanton of topological charge
\eqn\thirty{ p = {1 \over 8 \pi^2} \int \Tr F^{(4)} F^{(4)} . }
will give rise to $p$ chiral fermion zero modes on the string, each carrying
charge $q= 1/(2 \sqrt{3})$ under the unbroken  gauge group $G' \equiv U(1)$.

There are two fermionic contributions to the current, but unlike the axion
string model where their divergence added to zero, here the total covariant
divergence must  equal  the six-dimensional consistent anomaly.
The covariant divergence of the $U(1)$ component of the six-dimensional
consistent current, integrated
over the space transverse to the string is
\eqn\thhalf{\eqalign{ \int_{R^4} \Tr ( T^8 D J_f^\con ) =&  - c'  \int_{R^4}
                                 \Tr (T^8 \dtwo \Atwo)   \Tr  \Ffour \Ffour
\cr
                             = & - {c \over 24 \pi} \Tr ( T^8 \dtwo \Atwo ) p =
                            -{p \over 96 \pi} \dtwo \Atwo_8 . \cr  }}
Here  $T_8 = - i {\rm diag} ( 1, 1, -2)/2 \sqrt{3}$ is the $U(1)$ generator and
$A_8 = -2 \Tr T^8 A$ is the $U(1)$ component of the gauge field.  The two
fermion contributions to the current must sum to give \thhalf.

The contribution from the chiral zero modes is given by the two-dimensional
consistent anomaly and is
\eqn\ththird{ \Dtwo j_8^\con = {q^2 \over 4 \pi}  \dtwo \Atwo_8 p =
                        {p \over 48 \pi} \dtwo \Atwo_8 .}
We thus infer that there must be a contribution from the non-zero mode
fermions, $J_f^{\rm nz}$ which is not
purely two-dimensional and which satisfies
\eqn\thfourth{ \int_{R^4} \Tr(  T^8 D J_f^{\rm nz} )= - {p \over 96 \pi} \dtwo
\Atwo_8 .}
%
%

We now turn to the bosonic contribution to the current.  Since $D(J_b +
J_f^\con)$=0
it is clear that the total current is conserved, \ie\  that
\eqn\stupid{ \Dtwo j_8^\con + \int_{R^4}  \Tr T^8 D(J_b + J_f^{\rm nz}) = 0 }
but it is interesting to note that $J_b$  has physically distinct
contributions.
The first term in the bosonic current \twtwo\ falls off as $1/r^3$ in an
instanton
background and  gives rise to a radial current inflow at infinity.  The
integral from
this first term is
\eqn\fterm{ \int_{R^4} \Tr T^8 D (-2 \alpha'  \Ftwo {}^*H )  = - {p \over 48
\pi} \Ftwo_8.}
The second
term falls of much faster with $r$ and plays the role of the term $\Delta J$ in
the
axion string discussion which was needed to convert the consistent current
to the covariant current. In the axion string example it was a delta function
on
the string because we were working in the thin string approximation. Here it
is smeared over the whole core of the instanton.  The role of this term will be
clearer when we consider a non-abelian $G'$ since for $G'=U(1)$ the divergence
of the
consistent current is ``accidentally'' covariant.  The integral of this second
term is
\eqn\sterm{ \int_{R^4} \Tr T^8 D (\alpha'  \Atwo d {}^* H) =  {p \over 96 \pi}
\Dtwo \Atwo_8.}
The third term in $J_b$ is closed and does not contribute to the current
inflow.

As a second example which will bring out the role of the bosonic current terms
and also illustrate some of the group theoretical subtleties we take the
gauge group to be $G= E_6 \supset SU(6) \times SU(2) $
with fermions in the adjoint ${\bf 78}$ and the embedding given by
\eqn\esix{{\bf 78} \rightarrow ({\bf 35},{\bf 1}) + ({\bf 20},{\bf 2}) + ({\bf
1}, {\bf 3}). }
If we normalize the $E_6$ generators so that they obey the previous
normalization
for an $SU(2)$ subgroup  we find $\CC= 1/12$.  Since $E_6$ has no independent
quartic Casimir  \eighteen\ is satisfied and one can show  that
$c=1/32$ using the $SU(2)$ embedding \esix.  An $SU(2)$ instanton with
topological
charge $p$ gives rise to $|n|$ zero modes with
\eqn\nzmode{ n = {1 \over 8 \pi^2} \int_{R^4}  \Tr \Ffour \Ffour =24 p .}
{}From the decomposition \esix\ we see that $4p$ of these zero modes are
singlets
under $G'=SU(6)$ while the rest transform as p  ${\bf 20}$'s  of $SU(6)$.

Evaluating the total fermion contribution to the anomaly for a particular
$SU(6)$ current with generator $T^A$  as in \thhalf\   we now find
\eqn\thhagain{ \int_{R^4} \Tr( T^A D J_f^\con ) = - {c \over 24 \pi} \Tr ( T^A
\dtwo \Atwo )
24 p  = - {p \over 32 \pi} \Tr ( T^A \dtwo \Atwo ) . }
The contribution from the chiral zero modes is again given by the consistent
anomaly as
\eqn\freds{ \Tr T^A  \Dtwo j^\con = - { p \over 4 \pi}  {\rm Tr}_{\bf 20} T^A
\dtwo \Atwo .}
It is crucial to note that in \freds\ the trace on the right hand side is in
the  representation
of the chiral zero modes, that is the ${\bf 20}$
of $SU(6)$.  The traces in \thhagain\ and \freds\ can be related by noting
that for $SU(6)$,   $\Tr_{\bf 35} (T^A T^B) = 2 \Tr_{\bf 20} (T^A T^B)$.  Thus
the
two-dimensional contribution can be written as
\eqn\dumb{\Tr T^A \Dtwo j^\con =  - {p \over 16 \pi} \Tr ( T^A \dtwo \Atwo )}
and as before is twice the contribution \thhagain.  The fact that these factors
work out
in this way may at first sight seem miraculous. The total fermion contribution
\thhagain\
depends on the constant $c$ appearing in the factorization \eighteen\ while the
two-dimensional contribution depends on the embedding of $G'$ in $G$ and the
relative normalization of the two traces. Of course these two factors are not
independent as can be seen by  calculating $c$ by choosing the generators
to be in $G'$ and using the embedding of $G'$ in  $G$.   As before we
deduce that there must be an additional contribution from the non-zero mode
fermions which when added to \dumb\ gives the total fermion contribution
\thhagain:
\eqn\abc{\int_{R^4} \Tr T^A D J_f^{\rm nz} =  { p \over 32 \pi} \Tr T^A \dtwo
\Atwo . }

Turning now to the bosonic current there are again two terms that
contribute. The first is
\eqn\bcd{\eqalign{ -2 \alpha' \int_{R^4} \Tr T^A D( F {}^*H) =&  {2 c \over
(2\pi)^3 4! }
                   \Tr (T^A \Ftwo) \int_{R^4} \Tr \Ffour \Ffour \cr  =&  {p
\over 16 \pi}
                   \Tr( T^A \Ftwo) }}
while the second is
\eqn\cde{ \alpha' \int_{R^4} \Tr ( T^A D ( A d {}^* H)) = - {p \over 32 \pi}
\Tr T^A \Dtwo \Atwo .}
Adding together the contributions  \bcd,   \cde, \abc, and \dumb\ we find
\eqn\superdumb{ {p \over 16 \pi} \Tr T^A \left(  - \Ftwo + \half \Dtwo \Atwo
-\half \dtwo \Atwo
                              + \dtwo \Atwo \right) = 0 .}

Note that the first contribution to the bosonic current is covariant and gives
the inflow from infinity while the second term is of the from $ \Dtwo \Delta J$
found
in the axion string analysis.  We would also expect
the contribution
from the non-zero mode fermions to have two physically distinct
contributions.

It would  be nice to have a better understanding of the non-zero-mode
contribution to the fermion current.
A current $J_f^{\rm nz}$ which satisfies \abc\ or \thfourth\
can be obtained by varying the local action
\eqn\threethree{ S_{\rm nz} = c' \int \omega_3^{G'}  \omega_3^{SU(2)} }
where the superscripts indicate that the Chern-Simons terms are to be
evaluated only in the given subgroup. Varying \threethree\
gives
\eqn\jagain{J_f^{\rm nz} = c' ( 2 F^{G'} \omega_3^{SU(2)} - A^{G'} d
\omega_3^{SU(2)}  )}
which as mentioned above contains a covariant part which falls off as
$1/r^3$ and a non-covariant term localized on the string. However
the action \threethree\ cannot be obtained from a local $G$ invariant
action since $\omega_3^G \omega_3^G = 0$.  We can also understand
the action \threethree\ as follows. At large distances from the string the
$SU(2)$ gauge field approaches a pure gauge configuration.  We can thus remove
the coupling of the $SU(2)$ gauge field to the fermions by a gauge
transformation.
However, because of the anomaly in the fermion current given by \nineteen,
this induces an effective coupling between the $SU(2)$ and $G'$ gauge
fields given by \threethree. This argument only gives the action \threethree\
far from the string.

\newsec{Fivebranes in Ten Dimensions}

We now turn to an analysis of fivebrane solutions in ten dimensions.
Our analysis will parallel that in the previous section except that
now we will include both the gauge, gravitational, and mixed anomalies
in our analysis.

A number
of fivebrane solutions to string theory are known \fbranes.  We will consider
the
``gauge'' solution originally discovered by Strominger \hetsol.  The bosonic
action we start with is the bosonic sector of $N=1$ ten-dimensional
supergravity
coupled to $N=1$ super Yang-Mills with gauge group $G=SO(32)$ or
$E_8 \otimes E_8$.  This action is
\eqn\thfour{ S = \int  d^{10} x \sqrt{g} e^{-2 \phi} \left( R + 4 \partial_\mu
\phi
                               \partial^\mu \phi - {1 \over 3} H_{\mu \nu \rho}
H^{\mu \nu \rho} -
                    { \alpha'   \over 30} \Tr F_{\mu \nu} F^{\mu \nu} \right) .
}
where
\eqn\thfourdefH{H=dB-{\alpha '\over 30}\omega _3^Y+\alpha '\omega _3^L}
and $\omega _3^Y$($\omega _3^L$) is the Chern-Simons three-form
for the gauge field (spin connection).

The inclusion of fermion fields gives an anomalous theory which allows
for anomaly cancellation as discovered by Green and Schwarz. The anomaly
in this theory is determined by a twelve-form $I_{12}$ which factorizes
in the form
\eqn\thfive{ I_{12} = {1\over {(2\pi)^6 96}} X_4  X_8 }
with
\eqn\thfivea{ X_4=\Tr R^2 - {1 \over 30} \Tr F^2 }
and
\eqn\thsix{ X_8 = {1 \over 24} \Tr F^4 - {1 \over 7200} (\Tr F^2)^2 - {1 \over
240}
                           \Tr F^2 \Tr R^2 + {1 \over 8} \Tr R^4 + {1 \over 32}
(\Tr R^2)^2 .}
Traces for the curvature are in the fundamental representation of
$SO(9,1)$. For the forms appearing in the above
equations we define other forms by descent as
\eqn\thsixa{\eqalign{X^1_{4n-2}(\Lambda) & =\Tr (\Lambda
X^A_{4n-2}) \cr dX^1_{4n-2} & =\delta _{\Lambda}X_{4n-1} \cr
dX_{4n-1} & =X_{4n} \cr }}
The gauge anomaly derived from $I_{12}$ then has the form
\eqn\thsixb{\eqalign{ G_{10}(\Lambda)&= - \int
\Tr (\Lambda D J_{fA}) \cr
 &=   c' \int ({2\over 3} X_6^1(\Lambda)X_4 +{1\over 3}X_2^1(\Lambda)X_8) , \cr
}}
and includes the contribution of Majorana-Weyl fermions of the gauge group
where
\eqn\thsixdef{c'={1\over {96(2\pi)^5}} }
There is also a gravitational anomaly which includes a spin $3/2$
gravitino piece as well as the gauge group fermionic contribution
\eqn\thsixg{\eqalign{ G_{10}(\Theta)= & \int
d^{10}x\, \epsilon^{\mu}\nabla^{\nu}T_{f\, \mu\nu}\cr
 = &  c' \int ({2\over 3} X_6^1(\Theta)X_4 +{1\over 3}X_2^1(\Theta)X_8)  \cr}}
and results from a lack of invariance under coordinate transformations
\eqn\thsixdefg{x^{\mu}\rightarrow x^{\mu}+\epsilon ^{\mu}}
with $\Theta_{\mu\nu}=\partial_{\mu}\epsilon_{\nu}-
\partial_{\nu}\epsilon_{\mu}$.
If we consider local Lorentz transformations,  this anomaly has
the same form, but the non-conserved current is $J_{f R}$
instead of the energy-momentum tensor.

Using
the variation of B, $\delta B=-\alpha ' X_2 ^1$,
these anomalies are canceled by adding a counterterm to the action given by
\eqn\thseven{ \Delta S =c' \int ( {1\over {\alpha '}}
(B X_8) -{2\over 3}X_3 X_7).}
In the analysis that follows the dilaton field and its equation of motion
do not play any role. We therefore  set $\phi=0$ for simplicity.

In the general case($E_8\otimes E_8$ or $SO(32)$) we obtain the
following equations of motion for $B$ and $A$
\eqn\theight{\eqalign{d\, ^*H &={c'\over {4\alpha'}}X_8 \cr
{4\alpha ' \over 30} D\, ^*F & =J_{fA}+J_{BA} \cr}}
where
\eqn\theighta{\eqalign{J_{BA}&= {4\over 15}\alpha 'F\, ^*H-
{2\over 15}\alpha 'Ad\, ^*H\cr &-{{2c'}\over 45}FX_7+{c'\over 45}AX_8
-{2c'\over 3}Y_{A9} . \cr}}
We will not need the precise form of the nine-form $Y_{A9}$ but only
the fact that $D Y_{A9} = X_6^A X_4$.  We then can check that
\eqn\theightb{DJ_{BA}={c'\over 90}dAX_8-{2c'\over 3}X^A_6 X_4=-DJ_{fA} . }
%
%
%
%
%

Turning to gravitational anomalies, they can be handled similarly
to gauge anomalies by focusing on the special case of local
Lorentz transformations of the spin connection $\omega$.
We can define a current $J_R$ associated with these transformations
and demand its conservation.  The curvature two-form is
\eqn\theleven{R=d\omega+\omega^2.}
The above action should, thus, be invariant under gauge transformations
of $\omega$
\eqn\thtwelve{\delta_{\Theta}\omega=D\Theta.}
Varying the action with respect to $\omega$ generates the current
\eqn\ththirteen{\eqalign{J_R &=J_{fR}+J_{BR}\cr
J_{BR}&=-8\alpha' R\, ^*H+4\alpha'\omega d\, ^*H\cr
&+{4\over 3}c'RX_7-{2\over 3}c'\omega X_8\cr
&-{2\over 3}c' Y_{R9}+dBRY_{R4}\cr}}
where $DY_9^R=X_6^RX_4$ and $dY_{R4}=0$.  Using \thsixdef, \theight,
and \ththirteen, we
see that $DJ_R=0$.

Since the ten dimensional theory is anomaly free,  we can proceed as in
section three to examine current inflow for an instanton fivebrane
background($M^6\times R^4$).  Fields and derivatives are split between
$M^6$ and $R^4$ as discussed previously.  The instanton is
embedded in an $SU(2)$ subgroup of the first $E_8$
such that $ E_8\supset SU(2)\times E_7$
with the decomposition $496\rightarrow (3,1)+(1,133)+(2,56)$.
The fivebrane spin connection is an $SO(5,1)$ gauge
field, and the Lorentz group decomposition is $SO(9,1)\supset
 SO(5,1)\times SO(4)$ where $10\rightarrow (6,1)+(1,4)$.
The number of zero modes contributing to the fivebrane anomaly
is $|n|$ where
\eqn\thfourteen{ n = {1 \over 8 \pi^2} \int_{R^4}  \Tr \Ffour \Ffour =60 p }
with $p$ the $SU(2)$ topological charge.  All zero modes contribute to the
gravitational anomaly but only the $p$ $56$'s  of $E_7$ are involved
in the gauge anomaly. Since the zero modes are all singlets under
the second $E_8$ factor, we ignore it in what follows.

Substituting the instanton solution gives the following total
fermionic covariant divergence  on the fivebrane
for the $E_7$ and Lorentz currents
\eqn\foone{\eqalign{ \int_{R^4} \Tr( T^A_{E_7} D J_{f}^\con ) &=
{p \over 48 (2\pi)^3} \Tr ( T^A_{E_7} \dsix \Asix )
({-1\over 900}\Tr (\Fsix)^2 +{1\over 60}\Tr(\Rsix)^2)\cr
 \int_{R^4} \Tr( T_{SO(5,1)} D J_{fR}^{\con} ) &=
{p \over 48 (2\pi)^3} [\Tr ( T_{SO(5,1)} \dsix \omegasix )
({1\over 60}\Tr (\Fsix)^2\cr &-{5\over 24}\Tr(\Rsix)^2)-
{1\over 6}Y_{R6}].\cr}}
The six-forms  $Y_{R6}$($Y_{G'6}$) are determined by
descent from $\Tr R^4$ ($\Tr _{G'}F^4$).

The zero mode consistent anomaly  on the fivebrane
is  determined from
\eqn\ftwo{I_8={-1\over (2\pi)^4} \left[  p ( {1\over 24}\Tr F^4-{1\over 96}
\Tr F^2\Tr R^2 ) +{n\over 128}({1\over 45}\Tr R^4+{1\over 36}
(\Tr R^2)^2) \right] }
where $n=60p$ here and the gauge traces are in the $56$ of
$E_7$.  Using the following relations
\eqn\fthree{\eqalign{{\Tr} _{56}F^4&={1\over 24}({\Tr} _{56}F^2)^2\cr
{\Tr} _{133}F^2&=3{\Tr} _{56}F^2\cr}}
and multiplying $I_8$ by one-half to account for a Majorana
condition on the zero modes, one calculates that
\eqn\ffour {\eqalign{\Tr (T^A_{E_7}  \Dsix j^\con_{E_7})&=
{3\over 2}\int_{R^4} \Tr( T^A_{E_7} D J_{f}^\con )\cr
\Tr (T_{SO(5,1)}  \Dsix j^\con_R)&=
{3\over 2}\int_{R^4} \Tr( T^A_{SO(5,1)} D J_{fR}^\con )\cr}}

We again deduce that there is a  non-zero mode fermion
contribution
\eqn\ffive{\eqalign{J^{nz}_{f}&={c'\over 90} F X_7
-{c'\over 180} A  X_8-{c'\over 3}Y_{9}\cr
J^{nz}_{fR}&= - {c'\over 6}(2RX_7-\omega X_8+Y_{R9})\cr}}
so that
\eqn\fsix{\eqalign{\int_{R^4}DJ^{nz}_{f}&= -{1\over 3}\Dsix j^\con_{E_7}\cr
\int_{R^4}DJ^{nz}_{fR}&= -{1\over 3}\Dsix j^\con_R\cr}}
As in  the six-dimensional case there is also a bosonic contribution to
the inflow which consists of a covariant current at infinity and a
term localized on the fivebrane which converts the six-dimensional
consistent current into a covariant current which correctly matches
the inflow from infinity.

The $SO(32)$ theory can be analyzed in a similar fashion by
embedding the instanton  in the first $SU(2)$
of $SO(32)\supset SU(2)\times SU(2)\times SO(28)$ where
$496\rightarrow (3,1,1)+(1,3,1)+(1,1,378)+(2,2,28)$.  Note that
the embedding $SO(32) \supset SU(2) \times SO(29)$ used in \iz\  is
not the minimal embedding and corresponds to a superposition of
two minimal instantons \bcgw. The number
of zero modes is again $|n|$ with $n=60p$ so that there are
$p$ $2\times 28$'s of $SU(2)\times SO(28)$ and four singlets.
%
%
%
%
In either the $E_8 \times E_8$ or $SO(32)$ model
 we infer a non-zero mode anomaly that is $- 1/ 3$ times
the consistent anomaly.  We can derive, for either
choice of the group $G$, the non-zero mode
contribution from the following local action:
\eqn\almost{S_{nz}={c'\over 3}\int Y^{G'}_7 \omega _3^{SU(2)}}
where $dY^{G'}_7=(2\pi)^3 24\, I_8$, and $G'$ is the fivebrane gauge
group.

\newsec{Discussion}
Topological defects which are solutions to gauge invariant theories
but which have chiral fermion zero modes which lead to apparent
anomalies in the low-energy theory on the defect must have a mechanism
to cancel this anomaly by inflow from the outside world. For scalar defects
this inflow can be determined fairly directly from the coupling of fermion
fields to the defect and the background gauge fields. For gauge defects
we have seen that the situation is a bit more subtle and involves
contributions both from the fermion fields and from the additional
interactions of bosonic fields which are required for anomaly cancellation
in the underlying theory.  A weak point in our analysis is the contribution
from the non-zero mode fermions. It would be nice to have a more
direct derivation and understanding of the actions \threethree\ and
\almost\ and the corresponding contributions to the currents.

There is a close relation between anomaly
cancellation in spacetime, and the cancellation of sigma-model
anomalies in the world-brane theory of defects as was pointed out
for the heterotic string in \hw.
In \iz\ this relation is explored in some detail and it is argued that
it is possible to redefine the current so that there is no inflow from
infinity. However, as also pointed out in \iz, this redefinition is only
sensible for closed defects which can be enclosed in a sphere of
finite radius.  Since such configurations are topologically trivial,
non-static, and do not possess fermion zero modes (at least for smooth
configurations)  the implications of the analysis in \iz\ are not
completely clear to us.  The situation for singular (fundamental) solutions
seems different as these have fermion zero modes even when
topologically trivial.  In any event, for infinite defects with non-trivial
topology there is definitely a current inflow from infinity which is
needed to make physical sense of the anomaly in the low-energy
effective theory on the defect.

It is also possible to turn the inflow argument around
and to {\it deduce} the presence of fermion zero modes on a defect
given the behavior of the fields around the defect at large distances.
This applies to both solitonic and fundamental objects.  As an example
consider a macroscopic fundamental string \refs{\witt, \dh, \dghr}.  In the
ground state there are no gauge fields excited outside the string and
an $H$ field satisfying
\eqn\lasta{d {}^* H = g {}^* V}
with $g$ a coupling constant.  If we now turn on spacetime gauge  and/or
gravitational fields tangent to the string worldsheet there will be an inflow
given  the first terms in the currents $J_{BA}$ and $J_{BR}$:
\eqn\lastb{ J_{BA} = {1 \over 30} (8 \alpha'  F {}^* H - 4 \alpha' A d {}^* H)
\qquad
                     J_{BR} = ( -8 \alpha' R {}^* H + 4 \alpha' \omega d {}^* H
) . }
As discussed in \witt\ and \iz,  with the correct normalization of
\lasta\ this inflow precisely matches the
sigma-model anomaly of the heterotic string which is derived by descent
from the four-form
\eqn\lastc{ X_4 = - {1 \over 16 \pi^2} ( {1 \over 30} \Tr F^2 - \Tr R^2 ) }

In \ddp\ it was pointed out that $X_4$ agrees with the result for {\it
two-dimensional}
Yang-Mills and gravitational anomalies for a string with $32$ left-moving
fermions and $8$ right-moving fermions
\eqn\lastd{ I_4 = - {1 \over 16 \pi^2} ( \tr F^2 - {r \over 24} \Tr R^2 ) }
with $r=32-8=24$ and the gauge trace in the vector representation (for $E_8$
the trace
is in the vector representation of the $SO(16)$ subgroup of $E_8$),
This is apparently the field content
of the heterotic string.  It was further argued that it should be possible to
deduce
fivebrane anomalies starting from the form $X_8$ appearing in the factorization
\thfive. This argument was criticized (correctly we believe) in \iz. The
problem
is that there are no {\it two-dimensional} gravitational anomalies in the
heterotic
string \heta\ and the $32$ left-moving fermions do not couple to the space-time
spin connection and thus do not contribute at all to the sigma-model anomaly.
Thus the equivalence between $X_4$ and $I_4$ is in the words of \iz\ a
``curious fact''  which is not relevant to the anomaly cancellation.

The above brings out an apparent difference between fundamental and
solitonic strings or defects. The fermion zero modes of solitonic defects,
such as those discussed
in earlier sections, necessarily couple to the spacetime spin connection
because they inherit this coupling from the spacetime fields from which they
are
constructed. This is not the case for fundamental  strings in the usual
formulation
or presumably
for fundamental fivebranes if such objects exist.

A possible although rather speculative explanation for the curious  equivalence
between $I_4$ and $X_4$
would be provided by the existence of a smooth soliton string  solution with
the same zero mode structure and long-range fields as the heterotic string but
with zero-modes coupling universally to the spacetime spin-connection.
The solution proposed in \duffstr\ does not satisfy this criterion because
it has long-range gauge fields and a different zero-mode structure.

One could also speculate about such an equivalence for solitonic
and fundamental fivebranes \ddp. Here the situation is even murkier
because the soliton solution has long-range gauge fields and the
fermion zero modes only transform under a subgroup of the full gauge
group ( a $(28,2)$ of  $SO(28) \times SU(2)$ in the case of the minimal charge
fivebrane
with $SO(32)$ gauge group).  On the other hand one might expect the
ground state of a fundamental fivebrane to  have no long-range gauge
fields and to act only as a source for the six-form potential which appears
in the dual formulation of $d=10$ supergravity.  Furthermore the zero
mode content conjectured in \ddp\ for a fundamental fivebrane consists
of a $32$ of $SO(32)$ and does not agree with the zero mode
content of the know soliton fivebrane when restricted to a
$SO(28) \times SU(2)$ subgroup.  Thus while the inflow argument
gives definite information about how the zero modes of a fundamental
fivebrane would have to couple to the spacetime spin and gauge
connection, the relationship between these zero modes and those
of the known soliton solutions  remains  obscure.

\bigskip\centerline{\bf Acknowledgements}\nobreak
J. H. would like to thank C. Callan, S. Naculich and A. Strominger for
discussions.
This work was supported in part by the
by NSF
Grant No.~PHY90-00386.  J.H.\ also acknowledges the support of NSF PYI
Grant No.~PHY-9196117.

\listrefs
\end